\documentstyle[prl,aps,epsfig]{revtex}
\begin{document}

\draft
\flushbottom
\twocolumn[\hsize\textwidth\columnwidth\hsize
\csname@twocolumnfalse\endcsname

\title{The Singular Effect of Disorder on Electronic Transport\\
in Strong Coupling  Electron-Phonon Systems}

\author{Sanjeev Kumar and Pinaki Majumdar }

\address{ Harish-Chandra  Research Institute,\\
 Chhatnag Road, Jhusi, Allahabad 211 019, India }

\date{Apr 17, 2005}

\maketitle
\tightenlines
\widetext
\advance\leftskip by 57pt
\advance\rightskip by 57pt
\begin{abstract}

We solve the  disordered Holstein model in three dimensions considering the 
phonon variables to be classical. After mapping out the  phases  of the `clean'
strong coupling problem, we focus on the effect of disorder at strong 
electron-phonon (EP) coupling. The presence of even weak disorder 
$(i)$~enormously enhances the resistivity $(\rho)$ at $T=0$, simultaneously 
suppressing the density of states at the Fermi level, $(ii)$~ suppresses the 
temperature dependent increase of $\rho$, and $(iii)$~leads  to a regime with 
$d\rho/dT <0$.  We locate the origin of these anomalies  in the disorder induced 
tendency towards polaron formation, and the associated suppression in effective 
carrier density and mobility. These results, explicitly at `metallic' density,
are of direct  relevance to disordered EP materials like covalent semiconductors, 
the manganites, and to anomalous transport in the A-15 compounds.

\

\

\end{abstract}

]

\narrowtext
\tightenlines

The effect of electron-phonon (EP) interactions 
in weak coupling systems  is well understood and
can be captured within Boltzmann transport theory. 
However, there are
materials,  {\it e.g}, the A-15 compounds,
where the electron-phonon (EP)
coupling is large and the interplay of disorder and
EP interaction leads 
to several anomalous features 
\cite{a-15-test-1,a-15-wies,a-15-tsuei,neg-tcr-dynes}
in the resistivity, $\rho$, including a 
rapid rise in the residual resistivity at weak disorder, a
sharp reduction in the density of states (DOS) at the
Fermi level,  suppression 
of the temperature $(T)$ dependence of  $\rho$ and, sometimes,
a regime with $d\rho/dT <0$.
The conjunction of EP coupling and disorder has  
dramatic effects on transport
in covalent semiconductors as well \cite{emin-buss}, and also 
controls the resistivity and optical spectra in the
low $T$ spin polarised phase of the manganites
\cite{mang-dis-ref}.

These  transport anomalies, ranging across a wide  variety 
of materials, have a common  origin in the
interplay of 
polaronic tendency (arising from  strong EP coupling)
 with quenched disorder and thermal fluctuations.
However, to capture these effects within a unified framework
we need to handle the strong coupling and quenched disorder
non perturbatively, and simultaneously 
retain the spatial correlations in the 
problem.

We accomplish that in this paper, solving the 
Holstein model in three dimensions, with  
arbitrary coupling and quenched disorder, through a new real space
Monte Carlo (MC) technique \cite{tca-ref} 
explicitly retaining the spatial correlations
and localisation effects in the problem. We work with adiabatic phonons, 
{\it i.e},  treat the phonons as annealed classical variables. 
Apart from providing the first  complete solution to 
this   
problem, our results specifically highlight:
$(i)$~the huge amplification of quenched disorder
by strong EP coupling, and the dramatic increase in residual resistivity 
with weak disorder, $(ii)$~the rapid suppression of the `thermal'
component of $\rho(T)$ with increasing disorder, and $(iii)$~the 
emergence of a regime with $d\rho/dT <0$. We correlate these effects with
$(a)$~the appearance  of a weak pseudogap 
in the density of states and $(b)$~the 
suppression of the optical spectral
weight,  suggest a transport phenomenology for these
disordered EP systems, and compare our results with 
data on the disordered A-15 compounds.

{\it Model:}
The disordered Holstein ($d$-H) model with spinless fermions and
classical phonons is described by:
\begin{equation}
H = -t\sum_{\langle ij \rangle}  
c^{\dagger}_i c^{~}_j 
+  \sum_{i }(\epsilon_i - \mu) n_i  
- \lambda \sum_i n_i  x_i  
+ H_K 
\end{equation}

The $t$ are nearest neighbour hopping on a simple cubic lattice, $\epsilon_i$
is the quenched binary  disorder, with value $\pm \Delta$,
$\mu$ is the chemical potential,  $\lambda$ is the
EP interaction, coupling electron density $n_i$ 
to the local distortion $x_i$, and
the `restoring force' arises from 
$H_K = (K/2) \sum_i x_i^2 $. 
The parameters in the problem are $\lambda/t$, $\Delta/t$,   
electron density $n$, and temperature $T$.
We  measure energy, frequency, $T$, {\it  etc}, in
units of $t$, and use  $K=1$.

While the strong coupling problem with one electron has been widely studied,
 we do not know of controlled
approximations to handle  strong coupling at metallic densities 
\cite{aubry},
except within dynamical mean field theory (DMFT).
DMFT has been used in the adiabatic limit
\cite{mill-mull} to map out the 
Fermi liquid and polaronic insulator regime, and the
occurence of charge order (CO)  \cite{cdw-ref}.
The effect of strong EP coupling on $\rho(T)$,
including the effect of   
disorder, has also been explored~\cite{mill-sds}.

Our present understanding of 
strong coupling EP systems owe much to DMFT, but 
this approach  has limitations in the presence of disorder. 
$(i)$~Disorder and strong EP interactions reinforce each other
and generate a 
strong local correlation between $\epsilon_i$ and $x_i$,
inaccessible in the  statistically
homogeneous description of DMFT.  
$(ii)$~In such a system the conductivity can no longer be computed in 
terms of the single particle
self energy, as in DMFT, and  vertex corrections leading to 
localisation are crucial to recover the $d\rho/dT<0$ regime.

With the possibility of  
polaron formation, and the need to handle disorder and thermal fluctuations,
MC calculation is the only unbiased method 
for approaching the $d$-H problem.  However,
the cost of annealing the phonon variables, via the standard
exact diagonalisation based MC  (ED-MC),
 grows rapidly with system size. 
For a cube
with $N=L^3$, the ED-MC cost grows as $N^4$, allowing typically $L=4$ with current
resources. 
We instead use a  
 ``travelling cluster approximation''
(TCA) \cite{tca-ref} 
for annealing the phonons.  Since TCA avoids iterative diagonalisation
of the electron Hamiltonian, and estimates the  energy cost of MC moves based on a
smaller `cluster' Hamiltonian, we can access sizes upto $12^3$ with relative
ease. We use a $4^3$ cluster for the TCA based updates. 
Once the phonon  variables are equilibriated, we 
diagonalise the {\it full electron 
Hamiltonian} in the equilibrium configurations
to compute \cite{sk-pm-scr} electronic properties. 
We use $N=8^3$, checking with $10^3$ for size effects. Our resistivity is in
units of $\hbar a_0/(\pi e^2)$, where $a_0$ is the lattice spacing.
As a rough measure, $\rho \sim 100$ in our units is equivalent to the
Mott resistivity, about $2-3$~m$\Omega$cm.

Let us start with the phases.
In the absence of disorder, the finite $T$ phases can be
broadly classified as $(i)$~Fermi liquid (FL)
with  $d\rho/dT >0$, $(ii)$~a polaronic liquid (PL), 
 with `large' lattice distortions, no  
long range positional order, and 
$d\rho/dT <0$,
and $(iii)$~the charge 
ordered
insulator 
(COI) state. Unlike at $T=0$
\cite{latt-pol1},
all the phases have 
finite density of states  at the Fermi 
level.
There is no `phase boundary' between FL and PL 
and the distinction is only in terms of $d\rho/dT$.

For  $\lambda \lesssim 2.0$, Fig.1.$(a)$, the only feature
in the $n-T$ 
phase diagram is the occurence of CO at $n=0.5$, and
the coexistence region separating the CO phase 
from the  
FL. 
With increasing $\lambda$ the CO region extends 
down to $n \sim 0.35$ (a chessboard pattern with `vacancies' \cite{latt-pol1}),
while $T_{CO}$ at $n=0.5$ grows and then 
diminishes slowly beyond $\lambda \simeq 3.5$.
We will discuss the CO phase in detail   
\begin{center}

\vspace{.3cm}

\begin{figure}
\epsfxsize=6.2cm \epsfysize=5.1cm \epsfbox{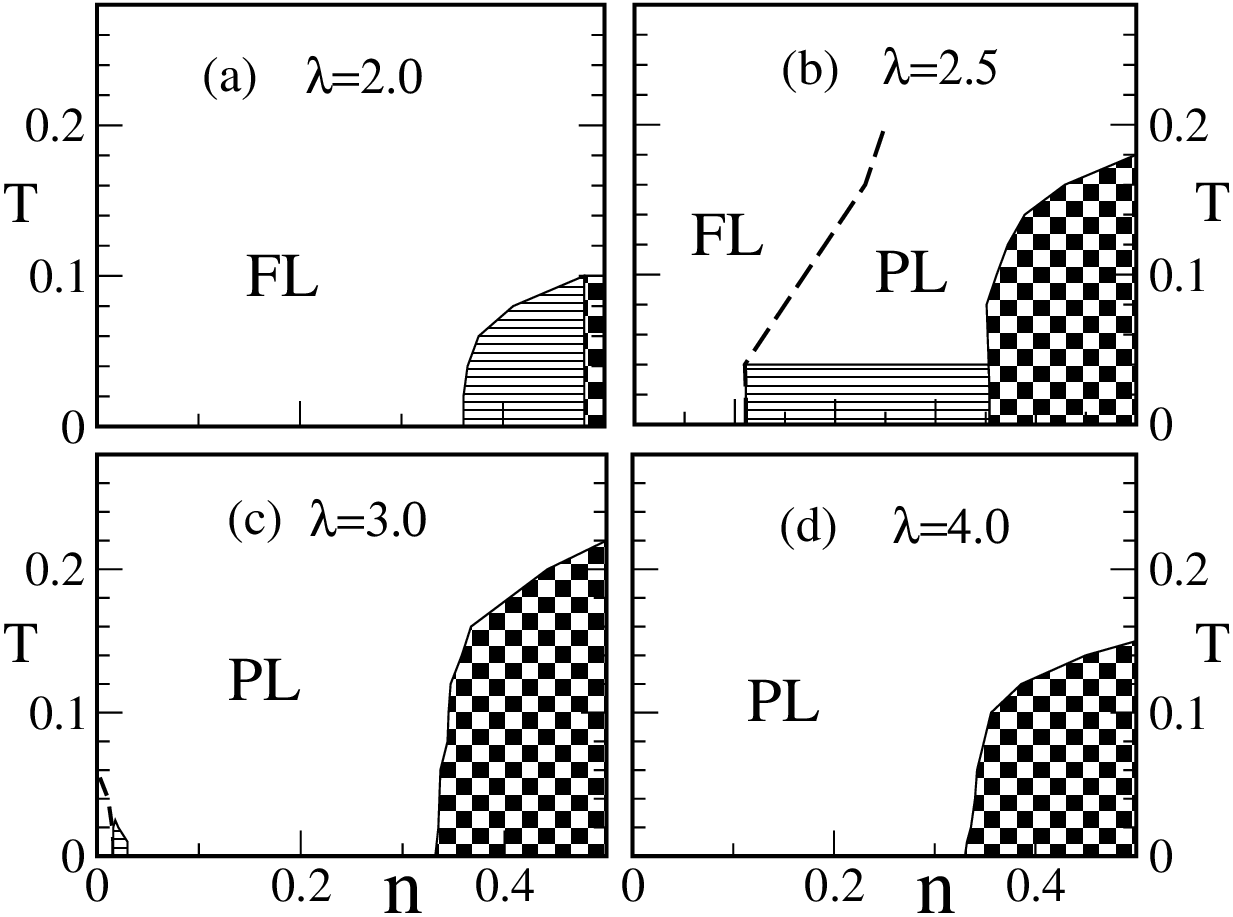}

\vspace{.3cm}

\caption{Phase diagram: clean model. Panels $(a)-(d).$ show the
$n-T$ phase diagram with increasing $\lambda$. The
``chessboard'' pattern represents a $\{ \pi, \pi, \pi \}$ CO phase, and the
hatched regions indicate phase separation. 
}
\end{figure}
\end{center}
elsewhere \cite{sk-pm-unpub}. 
The coexistence width at $T=0$ 
between the FL and COI increases with increasing $\lambda$ 
 but the temperature window for coexistence
reduces since the finite $T$ strong coupling FL 
itself has strong density fluctuations. 
By the time $\lambda =3.0$ the FL regime
has almost 
vanished,  the PL  and COI phases dominate the phase diagram, and
the system is insulating at all $n$ and $T$.

The interesting regime for studying the impact of disorder and
thermal fluctuations is roughly $\lambda \sim 2-3$ where the system
is at strong coupling but the metallic
state still survives. 
Our studies will be focused on $\lambda=2.0$, 
strong coupling but still metallic, and $n=0.3$,
and  we will use  $\lambda=1.0$ as 
our `weak coupling' reference.
To set the scale, the {\it single polaron} threshold is 
\cite{latt-pol1,romero}
$ \lambda_c = 3.3$, while at $n=0.3$ {\it collective} polaronic
localisation
\cite{latt-pol1} sets in  at $\lambda \approx  2.5$.
We study the
impact of disorder staying below this threshold.

The presence of disorder 
in the FL phase   tends to create density inhomogeneities,
$\delta n_{\bf r}$. This effect is  weak at weak disorder 
in an uncorrelated system but  
can be strongly amplified by a positive feedback
if the EP coupling is large. 
The phenomenon  is understood in the 
context of a single electron
\cite{emin-buss}, but an understanding of the dense 
many electron state, and
transport properties,  has  remained out of reach
and is our main focus.

The resistivity arises from a  
combination of quenched disorder and lattice distortions.
For a  
phonon configuration  $\{ x\}^{\alpha}$,
the ``potential'' seen  by the electrons
is  $\xi_i^{\alpha}  = \epsilon_i - \lambda x_i^{\alpha}$.
Subtracting  out the spatial average, $\bar {\xi }_{\alpha}$, 
the fluctuating part is
$ \delta \xi_i^{\alpha} \equiv \eta_i^{\alpha} = 
\epsilon_i - 
\lambda (x_i^{\alpha} - {\bar x_{\alpha}})$. 
Since the spatially averaged distortion does
not depend on configurations, {\it i.e},
${\bar x_{\alpha}} \approx {\bar x}$,
we can  write 
$ x_i^{\alpha} - {\bar x_{\alpha}} \approx  (x_i^0 -{\bar x}) + 
(x_i^{\alpha} - x_i^0)$, 
where  $x_i^0$ is the $T=0$ distortion at 
${\bf R}_i$. Let us 
define  $\delta x_i^0 \equiv  x_i^0 -{\bar x} $ and
$\delta x_i^{\alpha} \equiv x_i^{\alpha} - x_i^0 $, so that
$\delta x_i^0$ 
is the $T=0$ distortion at a site (with respect to the 
spatial average), and
$\delta x_i^{\alpha}$ is the thermal fluctuation {\it about that 
local distortion}. In terms of these, the fluctuating background
seen by the electrons is:  
$  \eta_i^{\alpha}  \sim  \epsilon_i - \lambda (\delta x_i^0
+ \delta x_i^{\alpha} )$. Averaging spatially, and over equilibrium 
configurations, the variance of this effective `disorder' is  
$ \eta^2 
= \langle (\epsilon_i - \lambda \delta x_i^0)^2 \rangle
 + \lambda^2 \langle (\delta x_i^{\alpha})^2 \rangle
- 2 \lambda \langle (\epsilon_i - \lambda \delta x_i^0)
\delta x_i^{\alpha} \rangle =
\eta_0^2 + \eta_T^2 + \eta_{corr}^2 $, say.
$\eta_0^2$ is a measure of disorder at $T=0$,
$\eta_T^2$ is a rough measure of ``thermal disorder'', and
$\eta^2_{corr}$ locally correlates the thermal fluctuations and 
the $T=0$
disorder.

In the clean  FL  there are no lattice distortions  at $T=0$,
{\it i.e},  $\delta x_i^0=0$, so
$  \eta^2    = \lambda^2 \langle (\delta x_i^{\alpha})^2 \rangle$. 
The stiffness to thermal fluctuations of the $x_i$, at low $T$
and weak coupling, can be
roughly estimated from the
energy cost 
$(K/2) x_i^2 + \lambda^2 \chi(n)x_i^2$,
where $\chi$ is the local, zero frequency,
density response function of the 
electron system. 
The effective stiffness is $K_{eff} = 
K + 2 \lambda^2 \chi(n)$
and   $\langle (\delta x_i^{\alpha})^2 \rangle \propto 
T/K_{eff}$. The thermal
disorder  is $ \eta^2  =\eta_T^2 \sim \lambda^2 T/K_{eff}$. 
Electron scattering from these fluctuations would lead to 
$\rho(T) \propto
\eta_T^2 $, for $\eta_T  \ll W$,  where $W=12t$ is the bare 
bandwidth,  but as $ \eta_T^2 $
 {\it grows}
the resistivity {\it rises faster than} $ \eta_T^2 $, due to
localisation corrections, unlike in DMFT.
The nature of $ \eta^2 $ and $\rho(T)$ in the clean limit 
and $\lambda=1.0$, and approximately at $\lambda=2.0$,  Fig.2,
can be understood within the above~scenario. 

For $\Delta \neq 0$ the effective disorder at  $T=0$ 
arises from $\epsilon_i$ {\it and} the induced lattice 
distortion: 
$ \eta^2 =
 \langle (\epsilon_i - \lambda \delta x_i^0 )^2 \rangle$.
If $\delta x_i^0$ is small, as one
expects at weak disorder and weak coupling, then $\eta^2
 \approx \langle \epsilon_i^2 \rangle =  \Delta^2$,
leading to $\rho(0) \sim \Delta^2$.
We have 
checked this at $\lambda=1$
and $n=0.3$, Fig.2.$(a)$ inset,
and the expected dependence clearly holds.
However, when $\lambda =2$,   
even weak disorder can create strong density inhomogeneities 
and induce {\it large } quenched 
distortions  \cite{emin-buss}. 
In this regime, $ \eta^2$ is dominated by
$\lambda^2 \langle (\delta x_i^0)^2 \rangle $. Since 
$\delta x_i^0 \sim (\lambda/K) \delta n_i^0$,
we have $ \eta^2 \approx (\lambda^4/K^2) (\delta n_i^0)^2$. 
Monte Carlo annealing of the phonon variables 
reveals that density inhomogeneities  can be strong, 
$\delta n_i^0 \sim {\cal O}(1)$, and the $\lambda^4$ factor leads
to a huge amplification of the external 
disorder.  
The rapid
growth in $\rho(0)$, Fig.2.$(c)$ inset,
arises due to this.
The small deviation from $\rho(0) \propto \Delta^2$, at small $\Delta^2$, 
is a finite size effect.

The effect of disorder and thermal fluctuations is additive 
in both $\eta^2$ and $\rho(T)$ at $\lambda=1.0$, as borne
out by the parallel curves in 
Fig.2.$(a)$ and Fig.2.$(c)$, {\it i.e}, $\rho(T)$ 
obeys Mathiessens rule.
However at $\lambda=2.0$ 
the same $\Delta$ leads to {\it much
stronger}  
effective disorder, $\eta^2(T)$, compared to
$\lambda=1.0$.
What is more striking is that despite 
the $d\eta^2/dT >0 $ in both cases  the ``thermal
increase''
in  $\rho(T)$ at $\lambda=2.0$ is actually {\it suppressed}
with increasing $\Delta$, Fig.2.$(d)$, leading eventually to a regime
with $d\rho/dT <0$, by the time  $\Delta \sim 0.5$.
We will argue that these transport anomalies at $\lambda=2.0$ arise
from the {\it disorder induced
polaron formation tendency,  
at $T=0$, and its weakening with
increasing temperature}. 
We next~discuss 
\begin{center}

\begin{figure}
\epsfxsize=7.0cm \epsfysize=6.8cm \epsfbox{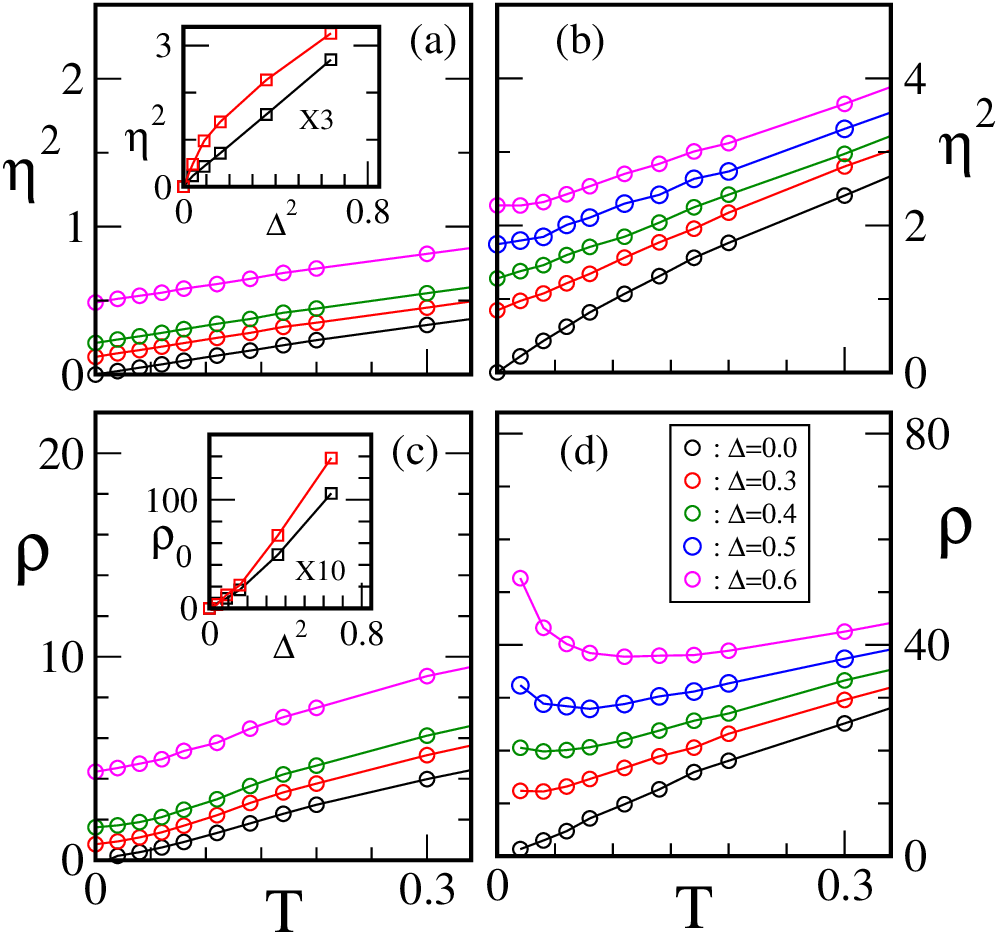}

\vspace{.2cm}

\caption{(Colour online) Effective disorder, $\eta^2$,  and resistivity, $\rho(T)$,
at $n=0.3$.
$(a)-(b).$  shows $\eta^2$: $(a).$~$\lambda=1.0$, 
$(b).$~$\lambda=2.0$.
$(c)-(d).$ $\rho(T)$, corresponding to the respective panel above.
Inset to $(a).$~$\eta^2(\Delta)$ at $T=0$,  black $\lambda=1.0$, red
$\lambda=2.0$, inset to $(c).$~$\rho(0,\Delta)$, black $\lambda=1.0$,
red $\lambda=2.0$.}
\end{figure}
\end{center}
\begin{center}
\begin{figure}

\vspace{.2cm}

\epsfxsize=7.0cm \epsfysize=7.0cm \epsfbox{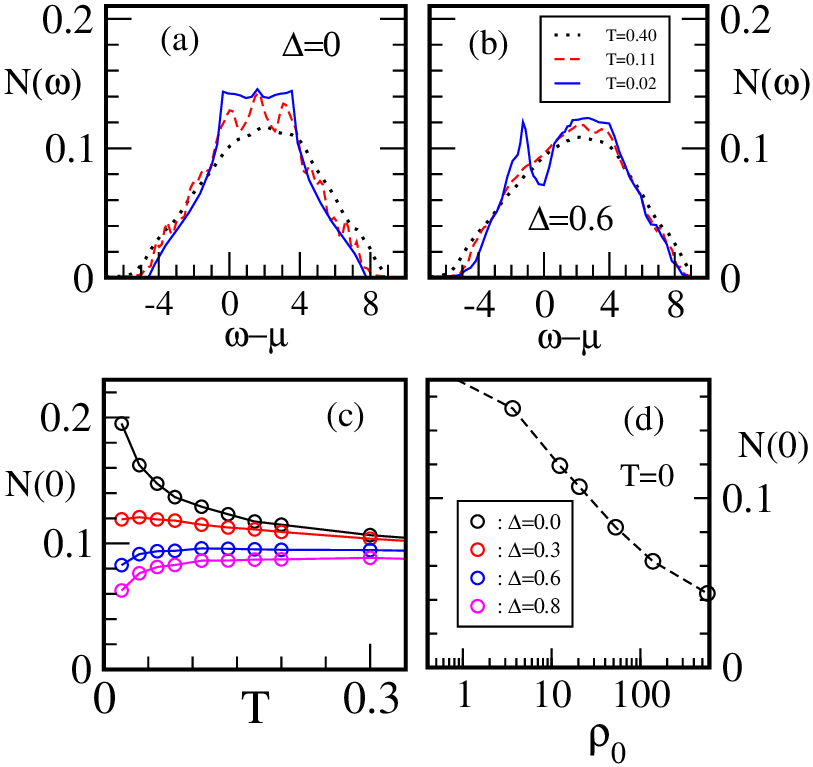}

\vspace{.2cm}

\caption{(Colour online) Density of states, at $\lambda=2.0$, $n=0.3$. 
The $\Delta$ and $T$ are marked on the panels.
$(c).$~Temperature dependence of $N(0)$, the
DOS at $\epsilon_F$, 
$(d).$~Correlation, at $T=0$,  between $N(0)$ and $\rho(0)$,
varying $\Delta$.
}
\end{figure}
\end{center}
the DOS and optics, and then our overall scenario.

The DOS, $N(\omega, T)$, is featureless in the clean problem, Fig.3.$(a)$,
even at $\lambda=2.0$, typical of gradual thermal disordering of a FL.
The presence of weak disorder, {\it e.g}, $\Delta=0.6$, Fig.3.$(b)$,
leads to the formation of a weak pseudogap around $\omega=0$.
The transfer of spectral weight to lower frequencies in $N(\omega)$ 
suggests possible localisation of some electrons into polaronic
 states. The pseudogap fills quickly with increasing $T$ and,
for $\Delta \lesssim 0.8$, vanishes for $T \gtrsim 0.15$, Fig.3.$(c)$.  
The suppression of DOS at $\epsilon_F$ and the increase in residual
resistivity, with increasing 
disorder, have a monotonic relation,
Fig.3.$(d)$.

The optical conductivity, $\sigma(\omega)$,
in the clean problem, Fig.4.$(a)$,
 is Drude like as expected of a Fermi liquid.
Very weak disorder, $\Delta=0.3$, Fig.4.$(b)$, retains
the Drude character, but suppresses the overall magnitude.
At $\Delta=0.6$, however, where anomalies were visible in
$\rho(T)$ and $N(\omega)$, the optical response is non Drude
at all $T$, 
suggesting the existence of localised states.
The  ``effective carrier number''
$n_{eff}({\bar \omega}, T)
= \int_0^{\bar \omega} \sigma(\omega, T) d\omega$, 
which  is a rough measure of the kinetic energy, 
is reduced by a factor $\sim 10$ (at ${\bar \omega }=1$) 
as $\Delta$ increases from zero
to $0.6$ at $T=0$, Fig.4.$(d)$.
With increasing $T$  the strongly
localised particles  delocalise, as borne out by the
increasing $n_{eff}(T)$, Fig.4.$(d).$,
and the DOS at $\epsilon_F$, Fig.3.$(b)$.

We propose the following  approximate ``two fluid'' framework
to  approach the results of Figs.2-4.
$({\bf a}).$~At  $T=0$ 
the clean adiabatic EP problem  allows only two kinds of
collective states \cite{latt-pol1} 
at $n=0.3$, $(i)$~a band metal
(with no lattice distortions) for $\lambda \lesssim  2.5$, 
or $(ii)$~a polaronic insulator, with all electrons localised, for
$\lambda \gtrsim 2.5$. The presence of weak disorder at
$\lambda =2.0$ induces strong 
localisation of {\it a fraction},
$f_{pol}=n_{pol}/n$ 
of charge carriers. 
$n_{pol}(\Delta) \approx \int_{x_{min}}^{\infty}
P(x, \Delta) dx$, where $P(x, \Delta)$ is the 
\begin{center}
\begin{figure}

\vspace{.2cm}

\epsfxsize=7.00cm \epsfysize=5.4cm \epsfbox{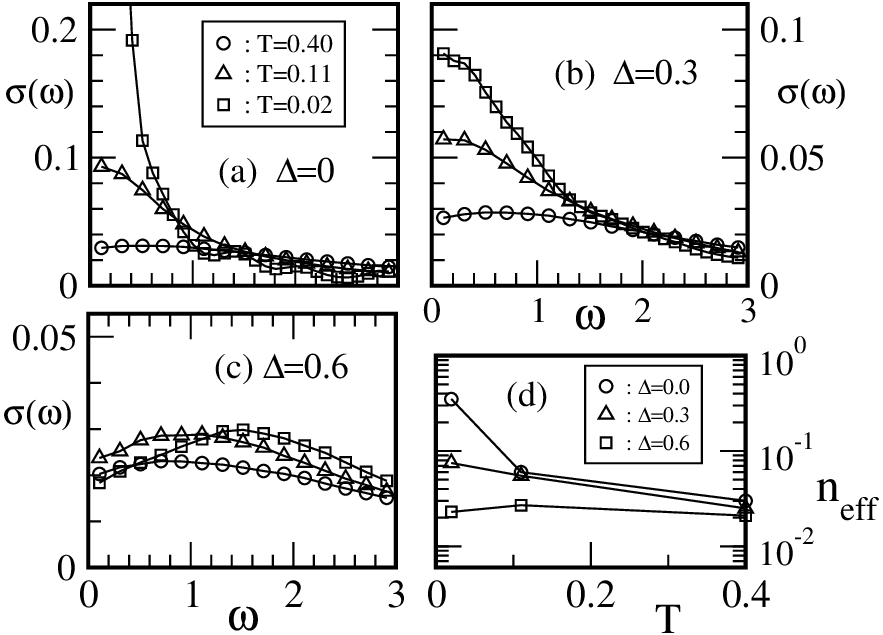}

\vspace{.2cm}

\caption{Optical response: $(a)-(c)$:
$\sigma(\omega, T)$ 
at $\lambda=2.0$ and $n=0.3$. $(a)$~$\Delta=0$, 
$(b)$~$\Delta=0.3$, $(c)$~$\Delta=0.6$. 
Panel $(d)$ $T$ dependence of $n_{eff}$, the integrated 
low frequency optical spectral weight (see text) for various $\Delta$.
}
\end{figure}
\end{center}
distribution of lattice distortions and
$x_{min}$ is the upper edge of the clean
FL peak.
However, states near $\epsilon_F$ 
continue to be delocalised. The fraction $f_{pol}$ 
increases with increasing  disorder \cite{sk-pm-unpub}, 
suppressing
$n_{eff}$ and $N(0)$ and increasing $\rho(0)$. 
The fraction, $1-f_{pol}$, of extended states are themselves
strongly scattered by the large distortions, $x_i$, but still
maintain `metallic' conduction.
$({\bf b}).$~With increasing $T$, the electrons in extended states
scatter off thermal fluctuations in $x_i$ and 
contribute to a growing $\rho(T)$,  while
localised states close below  $\epsilon_F$ 
provide a ``parallel'' conduction channel, 
whose {\it conductivity} increases with increasing  $T$. 
The reduction in $d\rho/dT$ at intermediate
$\Delta$ comes from these competing tendencies. 
For $\Delta \gtrsim 0.8$
there are few extended states and
at low $T$ the delocalising trend dominates.
We have explicitly checked the $T$ 
dependence of $n_{\bf r}$, \cite{sk-pm-unpub},
and the density inhomogeneity, due to strongly localised electrons,
 weakens with increasing $T$.
At higher $T$, after the polaronic effects have
disappeared,  $\rho(T)$ again roughly
tracks $\eta^2$. 

Some care is needed in applying our single band adiabatic 
results to real materials, where 
$(i)$~both the bandstructure and the
EP coupling could be more complicated, and $(ii)$~the phonon frequency,
$\omega_{ph}$, although usually $\ll t$ is still finite.
While the former will affect any quantitative comparison, the later 
brings in the physical effect of phonon assisted hopping, and 
possibly a larger conductivity from the `localised' states than
in the $\omega_{ph}=0$ limit. Such a $(\omega_{ph} \neq 0)$
mechanism had been studied 
in the 
limit of {\it strong} disorder and {\it weak} coupling \cite{girv-jon}
to explore the sign change in $d\rho/dT$.
Our focus is on the complementary {\it strong} coupling, {\it weak}
disorder~limit.

The 
interplay of disorder and strong EP coupling is best documented
in the `high T$_c$'  A-15 compounds
\cite{a-15-test-1,a-15-wies,a-15-tsuei,a-15-new}. There, in addition 
to `resistivity saturation' \cite{res-sat} in the clean limit,
the combination of strong EP coupling with 
weak disorder
leads to the following effects:
$(i)$~$\rho(0)$ increases  {\it enormously}, 
in response to disorder (alpha particle damage)  \cite{a-15-test-1}
compared to similar disorder in a `weak coupling' material, Nb, say, 
$(ii)$~as $\rho(0)$ grows,
the thermal increase in $\rho(T)$ is systematically reduced,
tending to a limit $d\rho/dT \sim 0$ at strong disorder \cite{a-15-test-1},  
$(iii)$~there is a degradation of $T_c$ with increasing $\rho(0)$,  
 suggesting a reduction in the
DOS at $\epsilon_F$ \cite{a-15-wies,a-15-tsuei}.   
$(iv)$~Some other systems, 
{\it e.g},  LuRh$_4$B$_4$ actually go over to $d\rho/dT <0$ with
increasing $\rho(0)$ \cite{neg-tcr-dynes}. 
{\it All the features, 
$(i) - (iv)$ above}, find a consistent explanation 
in terms of our results. The reduction in DOS inferred 
from $T_c$ degradation
bears a close parallel to our correlation between $N(0)$ and
$\rho(0)$,~Fig.3.$(d)$.

{\it Conclusion:}
In this paper we solved the problem of transport
in a dense, strong coupling, disordered electron-phonon system,
explicitly in three dimensions.
Our results reveal how disorder affects the many  electron system,
close to a collective polaronic instability, by strongly 
localising a {\it fraction}
of electronic states, and dramatically reducing the mobility of
extended states that survive near the Fermi level. We  mapped
out the interplay of these polaronic and extended states  
on transport, spectral, and optical properties, suggested a 
phenomenology for these class of systems, and compared our results
with data on the A-15 compounds.

We acknowledge use of the Beowulf cluster at H.R.I.

{}

\end{document}